\documentclass[twocolumn, prl]{revtex4}
\usepackage{graphicx}
\usepackage{epsfig}
\usepackage{color}
\usepackage{amsmath,bm}

\begin{document}
\title{$s$-wave Contacts of Quantum Gases in Quasi-one and Quasi-two Dimensions}
\author{Mingyuan He$^1$, Qi Zhou$^2$}
\affiliation{1. Department of Physics, The Chinese University of Hong Kong, Shatin, New Territories, Hong Kong\\
2. Department of Physics and Astronomy, Purdue University, West Lafayette, IN, 47907}
\date{\today}
\begin{abstract}

In quasi-one- or quasi-two-dimensional traps with strong transverse confinements, quantum gases behave like strictly one- or two-dimensional systems at large length scales. However, at short distance, the two-body scattering intrinsically has three-dimensional characteristics such that an exact description of any universal thermodynamic relation requires three-dimensional contacts, no matter how strong the confinement is. A fundamental question arises as to whether one- or two-dimensional contacts, which were originally defined for strictly one or two dimensions, are capable of describing quantum gases in quasi-one- or quasi-two-dimensional traps. Here, we point out an exact relation between the three- and low-dimensional contacts in these highly anisotropic traps. Such relation allows us to directly connect physical quantities at different length scales, and to characterise the quasi-one- or quasi-two-dimensional traps using universal thermodynamic relations that were derived for strict one or two dimensions.

\end{abstract}

\maketitle

A striking property of dilute quantum gases is that only a few physical quantities, the so-called contacts, fully govern a complex quantum many-body system. Contacts connect distinct physical observables through universal thermodynamic relations that are valid regardless of the microscopic parameters, and provide physicists a unique and powerful tool to bridge few-body and many-body physics. In the past decade, the study of contacts and universal thermodynamic relations has become a fundamentally important topic in quantum gases \cite{ Tan1,Tan2,Tan3,V1,V2,V3,V4,V5,Vale,Jin1,Jin2,Jin3,T1, T2,T3, T4,T5,Zhou,Drut} and attracted significant interest from nuclear physicists and other communities \cite{N1,N2,N3}. Whereas the original work on contact focused on the $s$-wave one \cite{Tan1,Tan2,Tan3}, recent studies have generalized such concept to high partial-wave contacts \cite{P1,P2,Zhou1,P3, P4,Cui}. It has also been realized that, to have a complete description of the universal thermodynamic relations, contacts should be defined as a matrix \cite{Zhou2,P5}. 

Similar to other physical quantities and phenomena, contacts and universal thermodynamic relations exhibit distinct behaviours in different dimensions \cite{V3,V4,V5}. For instance, the three-dimensional (3D) $s$-wave contact, $C_{3D}$, is proportional to $\frac{\partial E}{\partial (-1/a_{3D})}$ at the ground state, where $E$ is the total energy, and $a_{3D}$ is the 3D scattering length. In contrast, contacts in one dimension (1D) and two dimension (2D) are proportional to $\frac{\partial E}{\partial \ln(a_{2D})}$ and $\frac{\partial E}{\partial a_{1D}}$, where $a_{1D}$ and $a_{2D}$ are the scattering lengths in 1D and 2D, respectively. Other universal thermodynamic relations also have qualitative differences in different dimensions. The origin of such fundamental differences is the distinct asymptotic form of the wavefunctions near the origin, which behaves like $1/r$, $\ln \rho$, and $|z|$ in 3D, 2D and 1D, respectively, where $r$, $\rho$, $z$ are the coordinates in corresponding dimensions. Despite that drastic progress has been made in studying contacts in the past decade, works in the literature have been treating contacts in different dimensions separately. The dimension crossover, a class of problems of fundamental interest to both condensed matter and quantum gases communities, has not been considered for contacts and universal thermodynamic relations \cite{Olshanii1,Ketterle,Olshanii2,Qin,Petrov}.

\begin{figure}
	\centering
	{\includegraphics[width=0.48\textwidth]{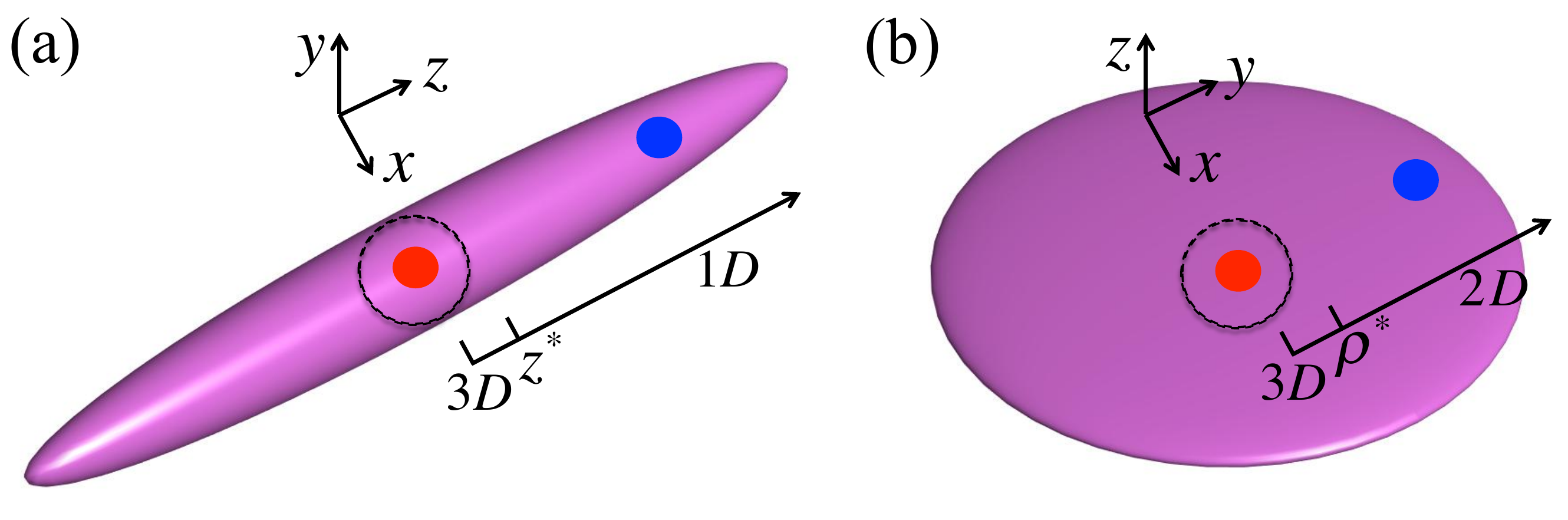}}
	\caption{(a) A quasi-1D trap. Atom cloud (purple cloud) with a strong harmonic confinement in the $x$-$y$ plane. Red and blue spheres represent a spin-up and spin-down atom, respectively. When their separation is much larger (smaller) than $z^{*}\sim d^{-1}$, two-body scatterings have 1D (3D ) features, and $C_{1D}$ ($C_{3D}$) controls all physical quantities in the corresponding large (small) length and small (large) momentum scales.  (b) A quasi-2D trap with a strong harmonic confinement along the $z$ direction. $C_{2D}$ ($C_{3D}$) controls the system in a scale $\rho\gg \rho^{*}\sim d^{-1}$ ($\rho\ll \rho^{*}$). }\label{Fig1}
\end{figure}

In laboratories, a 1D or 2D system is created by applying a tight confinement, for instance, a strong harmonic trap of a harmonic oscillator length $d$ and frequency $\omega$, along one or two spatial directions, as shown in figure \ref{Fig1}. Such systems are often referred to as quasi-1D or quasi-2D systems. When the characteristic many-body energy scales, for instance, the chemical potential, are much smaller than $\hbar\omega$, it is well known that the system behaves like a strictly 1D or 2D system if the long-wavelength or low-energy physics is considered. However, in a length scale much smaller than $d$, the two-body interaction has essentially 3D characteristics, as the confining potential can barely affect the two-body wavefunction in such regime. The asymptotic form of the many-body wavefunction derived for a strictly 1D or 2D system then does not apply to quasi-1D or quasi-1D traps in the zero range limit of the distance between two atoms. Instead, the asymptotic form of a 3D many-body wavefunction applies, and $C_{3D}$ is required to describe universal thermodynamic relations in quasi-1D and quasi-2D traps, no matter how strong the transverse confinement is. Then a conceptual question arises, whether universal relations originally derived for strictly 1D and 2D systems still serve as exact descriptions of quasi-1D and quasi-2D traps?  To answer this question, it is desired to explore a number of fundamental questions, including the relations between $C_{3D}$ and $C_{1D}$ (or $C_{2D}$), how these contacts govern the quasi-1D (quasi-2D) traps in different length scales, and how universal relations in 3D transform to those in low dimensions.  

In this Letter, we establish an exact relation between $C_{1D}$ ($C_{2D}$) and $C_{3D}$ in quasi-1D (quasi-2D) traps. We focus on quantum gases with zero-range interactions such that only $s$-wave scatterings and $s$-wave contacts are relevant. Remarkably, $C_{1D}$ and $C_{2D}$ are related to $C_{3D}$ by simple geometric factors,
\begin{eqnarray}
&C_{3D}&=\pi d^2  C_{1D}, \label{C13}\\
&C_{3D}&=\sqrt{\pi d^2}  C_{2D}. \label{C23}
\end{eqnarray}
We will also show that $C_{1D}$ ($C_{2D}$) controls the momentum distribution $n({\bf k})$ in the intermediate momentum regime, $k_F\ll k\ll d^{-1}$, where $k_F$ is the Fermi momentum, $k=|{\bf k}|$, and $C_{3D}$ determines $n({\bf k})$ in the large momentum regime, $k\gg d^{-1}$. Though the same physical quantity in a given length or energy scale is often uncorrelated to that in a distinct scale, here, Eq. (\ref{C13}) and Eq. (\ref{C23}) allow one to establish the exact relations of $n({\bf k})$, as well as any other physical quantities, in distinct scales, a unique property of quantum gases in quasi-1D and quasi-2D traps. Eq. (\ref{C13}) and Eq. (\ref{C23}) also provide us an unprecedented means to explore universal thermodynamic relations using two equivalent schemes, i.e., either through  $C_{3D}$ that controls any physical systems, including highly anisotropic traps, or using $C_{1D}$ ($C_{2D}$), which was originally defined in strict 1D (2D) and governs the long-range physics in the quasi-1D (quasi-2D) traps.  We will illuminate this fact by using the adiabatic relation as an example.

We first consider a two-component fermion gases with total numbers $N_{\uparrow}$ and $N_{\downarrow}$ in each component in a quasi-1D trap. The Hamiltonian is written as
\begin{equation}
H =  - \sum\limits_i {\frac{{{\hbar ^2}\nabla _i^2}}{{2M}}}  + \sum\limits_i V ({\rho _i}) + g\sum\limits_{i = 1}^{{N_ \uparrow }} {\sum\limits_{j = {N_ \uparrow } + 1}^{{N_ \uparrow } + {N_ \downarrow }} {\delta ({{\bf{r}}_{ij}})\frac{{\partial \left( {{r_{ij}} \cdot } \right)}}{{\partial {r_{ij}}}}} } ,
\end{equation}
where $M$ is the mass of each atom, ${\bf r}_i=({\bm \rho}_i, z_i)$, $\rho_i=|{\bm \rho}_i|$, ${\bf r}_{ij}={\bf r}_i-{\bf r}_j$, $r_{ij}=|{\bf r}_{ij}|$, $V(\rho_i)=\frac{1}{2}M\omega^2\rho^2_i$ is a harmonic trapping potential for the $i$th atom in the $x$-$y$ plane. Atoms are free along the $z$ direction. $g=4\pi \hbar^2 a_{3D}/M$ is the strength of the Huang-Yang pseudopotential. $V(\rho_i)$ is sufficiently strong such that $d=\sqrt{2\hbar/(M\omega)}\ll k_F^{-1}$ is satisfied. This is equivalent to say that the chemical potential $\mu$ is much smaller than $2\hbar\omega$, the energy separation between the ground and the first vibration level of the harmonic trap.  When the distance between a spin-up and spin-down atom, which is denoted by $r=|{\bf r}|$, ${\bf r}={\bf r}_1-{\bf r}_2$, is much smaller than $k_F^{-1}$, the wavefunction of a many-body eigenstate has a universal asymptotic form
 \begin{equation}
\Psi \stackrel{r\ll k_F^{-1}}{\xrightarrow{\hspace*{1cm}} }  \int d\epsilon_q \phi({\bf r};\epsilon_q)G(\frac{{\bf r}_1+{\bf r}_2}{2}, {\bf r}_{i\neq 1,2}; \sigma_i;E-\epsilon_q) \label{Asym}
\end{equation}
where $\phi({\bf r};\epsilon_q)$ is the wavefunction of the relative motion of two atoms,  $\epsilon_q=\hbar \omega + \hbar^2q^2/M$ is the colliding energy, $q$ is the corresponding momentum, and $E$ is the total energy of the system.  $ {\bf r}_{i}$ and  $\sigma_i$ are the spatial and spin coordinates of the $i$th atom, respectively. Whereas Eq. (\ref{Asym}) is valid for any 3D systems, it is useful to make use of the explicit form of $\phi({\bf r};\epsilon_q)$ in quasi-1D traps,
\begin{equation}
\begin{split}
\phi({\bf r};\epsilon_q) =& \Phi_{00}({\bm \rho})[\cos(qz)+f(q)e^{iq|z|}]\\
&-f(q)\sum_{n>0}\frac{iq}{q_n}\Phi_{n0}({\bm \rho})e^{-q_n |z|},\label{2bwf}
\end{split}
\end{equation} 
where $\Phi_{nm}({\bm \rho})$ is the eigenstate of the harmonic trap with eigenenergy $E_\bot^{nm}=\hbar\omega(2n+|m|+1)$ in the $x$-$y$ plane, $n$ is the quantum number for the radial part of the wavefunction, and  $m$ is the angular momentum quantum number. $f(q)=i/[\cot{\eta_{1D}(q)}-i]$ is the scattering amplitude and $\eta_{1D}(q)$ is the phase shift in 1D. The first line in Eq. (\ref{2bwf}) is the contribution from the ground state of the harmonic trap, the second line is the contribution from excited states, and $q_n=\sqrt{(E_\bot^{n0}-\epsilon_q)M/\hbar^2}$. For $s$-wave scatterings, only wavefunctions with $m=0$ are relevant. Since $\hbar^2q^2/M$ is typically of the order of $\mu\ll 2\hbar\omega$, $q_n$ is positive for all $n>0$. Thus, the second line in Eq. (\ref{2bwf}) decays exponentially. When $|z|\gg z^*\equiv 1/q_1$, Eq. (\ref{2bwf}) reduces to a wavefunction in strict 1D. It is also easy to see that $z^*\sim d\ll k_F^{-1}$. Correspondingly, we obtain the momentum distribution in the regime $k_F\ll k\ll d^{-1}$, 
\begin{equation}
n_\sigma({\bf k}) \stackrel{k_F\ll k\ll  d^{-1}}{\xrightarrow{\hspace*{1cm}} } |\Phi_{00}({\bf k}_\bot)|^2\frac{ C_{1D}}{k_z^{4}}, \quad \sigma=\uparrow,\downarrow \label{nk1Dfull}
\end{equation}
where ${\bf k}=({\bf k}_\bot, k_z) $, $\Phi_{00}({\bf k}_\bot)=\int d^2{\bm \rho} \Phi_{00}({\bm \rho}) e^{-i {\bf k_\bot} \cdot {\bm \rho}}$, $\int d^2 {\bf k}_\bot |\Phi_{00}({\bf k}_\bot)|^2=(2\pi)^2$,
\begin{equation}
C_{1D}=4N_{\uparrow}N_{\downarrow}\int d^3{\bf R}_{12}\Big|\int d\epsilon_q qf(q)G({\bf R}_{12};E-\epsilon_q)\Big|^2,\label{C1D}
\end{equation}
and ${\bf R}_{12}$ is a short-hand notation for a set of coordinates $\{({{\bf r}_1+{\bf r}_2})/{2}, {\bf r}_{i\neq 1,2}; \sigma_i\}$, $ d^3{\bf R}_{12}=\prod\nolimits_{i \ne 1,2} {{d^3}{{\bf r}_i}} {d^3}\left( {{{\bf r}_1} + {{\bf r}_2}} \right)/2$. In this regime, $n_\sigma({\bf k})$ is a broad distribution along the $k_x$ and $k_y$ directions, as expected for a quasi-1D system. For $k_F\ll k_z\ll d^{-1}$, the expression in Eq. (\ref{nk1Dfull}) could be extend to $k_{\bot}\rightarrow \infty$, and $n_{1D}^\sigma(k_z)=\int \frac{d^2{\bf k}_\bot}{(2\pi)^2} n_\sigma({\bf k})$ is defined. We obtain
\begin{equation}
n_{1D}^\sigma (k_z)\stackrel{k_F\ll k_z\ll d^{-1}}{\xrightarrow{\hspace*{1cm}} } \frac{C_{1D}}{k_z^4},\label{nk1D}
\end{equation}
which recovers the result of a strictly 1D system.

We now consider $r\ll d$, where we have
\begin{equation}
\Psi \stackrel{ r\ll d}{\xrightarrow{\hspace*{1cm}} }(\frac{1}{r}-\frac{1}{a_{3D}}) \int d\epsilon_q G_{3D}({\bf R}_{12};E-\epsilon_q) \label{Asym3D}.
\end{equation}
Correspondingly, $n_\sigma({\bf k})$ has a large momentum tail,
\begin{equation}
n_\sigma({\bf k})\stackrel{k\gg d^{-1}}{\xrightarrow{\hspace*{1cm}} }\frac{C_{3D}}{k^4},\label{nk3D}
\end{equation}
where 
\begin{equation}
C_{3D}=(4\pi)^2 N_{\uparrow}N_{\downarrow}\int d^3 {\bf R}_{12} \Big|\int d\epsilon_q G_{3D}({\bf R}_{12};E-\epsilon_q)\Big|^2.\label{C3D}
\end{equation}
Indeed, Eq. (\ref{2bwf}) becomes $\frac{-iqf(q)}{2}\frac{d}{\sqrt{\pi}}(\frac{1}{|z|}-\frac{1}{a_{3D}})$ when $|z| \ll d$ for $\rho=0$, and \cite{Olshanii1}
\begin{equation}
a_{1D}=-\frac{d^2}{2a_{3D}} \left( 1-1.4603 \frac{a_{3D}}{d} \right) \label{sl13}
\end{equation}
where $\cot{\eta(q)}/q=a_{1D}$ and $G_{3D}({\bf R}_{12};E-\epsilon_q)=\frac{-iqf(q)}{2}\frac{d}{\sqrt{\pi}} G({\bf R}_{12};E-\epsilon_q)$. Compare Eq. (\ref{C1D}) and Eq. (\ref{C3D}), we immediately see that Eq. (\ref{C13}) holds. 

\begin{figure}
	\centering
	{\includegraphics[width=0.238\textwidth]{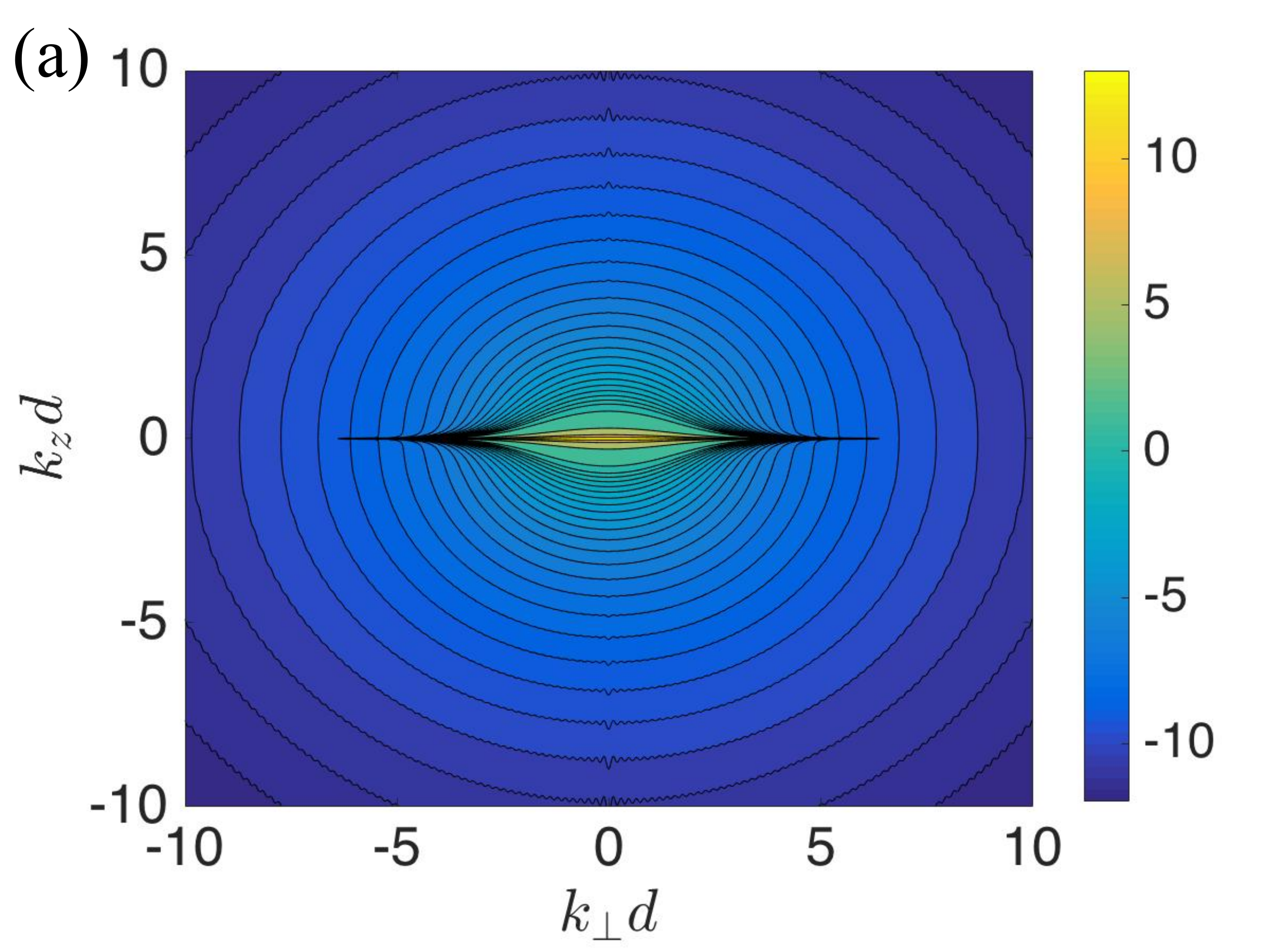}}
	{\includegraphics[width=0.238\textwidth]{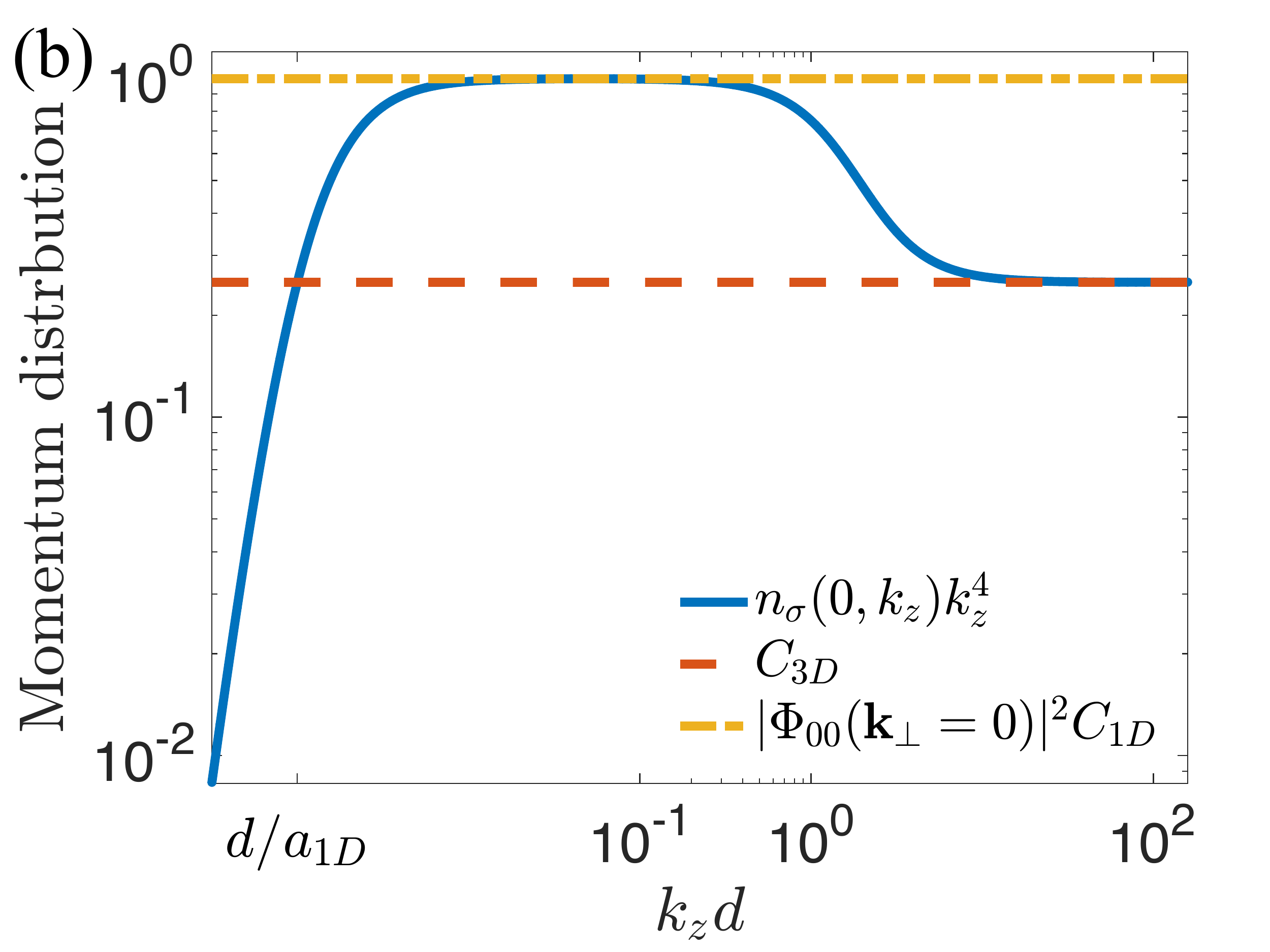}}
	\caption{(a) A contour plot of the exact momentum distribution $\ln(n_\sigma({\bf k}))$ of a two-body system, with $n_\sigma({\bf k})$ in unit of $d^4|\Phi_{00}({\bf k}_\bot=0)|^2 C_{1D}$. The total number of vibration levels considered is $N=300$, and $a_{1D}=1000d$. (b) Scaled momentum $n_\sigma(0,k_z)k_z^4$. It is determined by $C_{1D}$ and $C_{3D}$ in the regime $a_{1D}^{-1}\ll k_z\ll d^{-1}$ and $k_z\gg d^{-1}$, respectively. }\label{Fig2}
\end{figure}

It is interesting to note that Eq. (\ref{C13}) has a simple geometric interpretation. Though the quasi-1D trap is highly non-uniform in the transverse directions, it can be viewed as a cylinder with a uniform distribution of contact density on the cross section of radius $d$. Since the total contact in 3D is the contact density multiplied by the total volume, one can view  $C_{1D}$ as the linear contact density. Thus, $C_{3D}$ is simply $C_{1D}$ multiplied by the cross-sectional area $\pi d^2$. Eq. (\ref{C13}) also allows one to establish an exact relation between $n_\sigma({\bf k})$ in different momentum scales. From Eq. (\ref{nk1D}) and Eq. (\ref{nk3D}), we obtain
\begin{equation}
\begin{split}
n_\sigma({\bf k})k^4\big|_{k\gg d^{-1}}=(\pi d^2)n_{1D}^\sigma(k_z){k_z^4}\big|_{k_F\ll k_z\ll d^{-1}},
\end{split}
\end{equation}
a unique result originated from the exact relation between $C_{3D}$ and $C_{1D}$. 

To verify the above results, we evaluate exactly $n_\sigma({\bf k})$ of a two-body system using Eq. (\ref{Asym}) and Eq. (\ref{2bwf}). Its scaling behaviours also describe those of $n_\sigma({\bf k})$ in a generic many-body system in the regime $k\ll k_F$. By taking into account a large enough number of excited states, we obtain numerically $n_\sigma ({\bf k})$, as shown in figure \ref{Fig2}(a). Indeed, in the regime $k_F\ll k\ll d^{-1}$, $n_\sigma({\bf k})$ decays slowly with increasing $k_x$ and $k_y$. As aforementioned, the width of the wavefunction $\phi_{00}({\bf k_\bot})$ is given by the inverse of the harmonic oscillator length. Thus, for a strong confinement, $n_\sigma({\bf k})$ exhibits 1D feature in such momentum scale. In contrast, in the regime $k\gg d^{-1}$, $n_\sigma({\bf k})$ becomes isotropic, a 3D characteristic as expected. Figure \ref{Fig2}(b) shows the scaled momentum distribution $n_\sigma({\bf k})k^4$, which clearly demonstrates how $n_\sigma(0,k_z)$ gradually changes from $|\Phi_{00}({\bf k}_\bot=0)|^2{ C_{1D}}/{k_z^{4}}$ to ${ C_{3D}}/{k_z^{4}}$.

Besides $n_\sigma({\bf k})$, Eq. (\ref{C13}) allows us to connect other universal thermodynamic relations in 1D and 3D.  Here, we focus on the adiabatic relations. In strictly 1D systems, where the transverse degrees of freedom are absent, the adiabatic relation is written as \cite{V5}
\begin{equation}
\frac{dE}{da_{1D}}=\frac{\hbar^2C_{1D}}{2M}. \label{ab1D}
\end{equation}
In quasi-1D systems, as aforementioned, $C_{1D}$ controls physical quantities in a large length scale $z\gg d$, or equivalently, in the momentum scale $k\ll d^{-1}$. A complete description of the system needs the introduction of $C_{3D}$ to capture physics in the length scale $z<d$, or momentum scale $k>d^{-1}$. A natural question is then, whether Eq. (\ref{ab1D}) is still valid. 

Interestingly, a simple calculation shows that, Eq. (\ref{ab1D}) holds for quasi-1D system. The reason is that, Eq. (\ref{C13}) provides an exact relation between $C_{1D}$ and $C_{3D}$, the latter of which governs any 3D system, including a quasi-1D trap that is highly anisotropic. Thus the 3D adiabatic relation \cite{Tan2}
\begin{equation}
\frac{dE}{d(-1/a_{3D})}=\frac{\hbar^2C_{3D}}{4\pi M },\label{ab3D}
\end{equation}
is always valid in a quasi-1D trap. It is also known that $a_{3D}$ and $a_{1D}$ are related by Eq. (\ref{sl13}). Substitute this expression and Eq. (\ref{C13}) to Eq. (\ref{ab3D}), Eq. (\ref{ab1D}) is obtained. This immediately tells us that the adiabatic relation derived for strictly 1D systems applies to quasi-1D traps. In practice, Eq. (\ref{nk1D}) and Eq. (\ref{ab1D}) are also particularly useful, as experimentalists do not need to extract $C_{3D}$ from $n({\bf k})$ in the very large momentum regime $k\gg d^{-1}$, which may become too small to detect. Instead, a measurement of $n({\bf k})$ in the intermediate regime $k_F\ll k\ll d^{-1}$, which has a much larger amplitude, is sufficient to obtain $C_{1D}$ that could also fully governs the quasi-1D trap.

Whereas we focus on the adiabatic relation here, discussions can be directly generalised to other universal thermodynamic relations. Eq. (\ref{C13}) shows that any universal thermodynamic relations established by $C_{3D}$ can be rewritten in terms of $C_{1D}$ that governs the behaviours of the quasi-1D systems in the large length scale $z\gg d$. Thus, universal thermodynamic relations in 3D can be directly transformed to those in 1D.

We now turn to a quasi-2D trap. The Hamiltonian is written as
\begin{equation}
H =  - \sum\limits_i {\frac{{{\hbar ^2}\nabla _i^2}}{{2M}}}  + \sum\limits_i V (z_i) + g\sum\limits_{i = 1}^{{N_ \uparrow }} {\sum\limits_{j = {N_ \uparrow } + 1}^{{N_ \uparrow } + {N_ \downarrow }} {\delta ({{\bf{r}}_{ij}})\frac{{\partial \left( {{r_{ij}} \cdot } \right)}}{{\partial {r_{ij}}}}} } ,
\end{equation}
where $V(z_i)=\frac{1}{2} M \omega^2 z^2_i$ is a harmonic trapping potential for the $i$th atom along the $z$ direction. The system is free in the $x$-$y$ plane. The discussions are essentially parallel to those in 1D. Starting from Eq. (\ref{Asym}) and the two-body wavefunction in a quasi-2D trap for $s$-wave scattering, 
\begin{equation}
\begin{split}
\phi({\bf r};\epsilon_q) = & \frac{\pi}{2} \cot{\eta_{2D}(q) }[J_0(q\rho)-\tan{\eta_{2D}(q)} N_0(q\rho)]\Phi_{0}(z)\\
&+\frac{i\pi}{2} \sum_{n>0}(-1)^n \sqrt{ \frac{(2n-1)!!}{(2n)!!} }\Phi_{2n}(z)H_0^{(1)}(iq_{n}\rho), \label{2bwf2D}
\end{split} 
\end{equation}
it is straightforward to derive Eq. (\ref{C23}), the tails of the momentum distribution and the adiabatic relation. In Eq. (\ref{2bwf2D}), $\eta_{2D}(q)$ is the 2D phase shift, $J_0$ ($N_0$) is the Bessel function of the first (second) kind, $H_0^{(1)}$ is the Hankel function of the first kind, $\Phi_{n}(z)$ is the eigenfunction of harmonic oscillator along $z$-axis with eigen energy $E_z^n=\hbar\omega(n+1/2)$, $\epsilon_q=\hbar \omega/2 +\hbar^2q^2/M$ and $q_n= \sqrt{ (E_z^{2n} -\epsilon_q )M/\hbar^2}$. When $\rho>\rho^*\equiv 1/q_1$ ($\rho<\rho^*$), the wavefunction in Eq. (\ref{2bwf2D}) is 2D-like (3D-like).

\begin{figure}
	\centering
	{\includegraphics[width=0.238\textwidth]{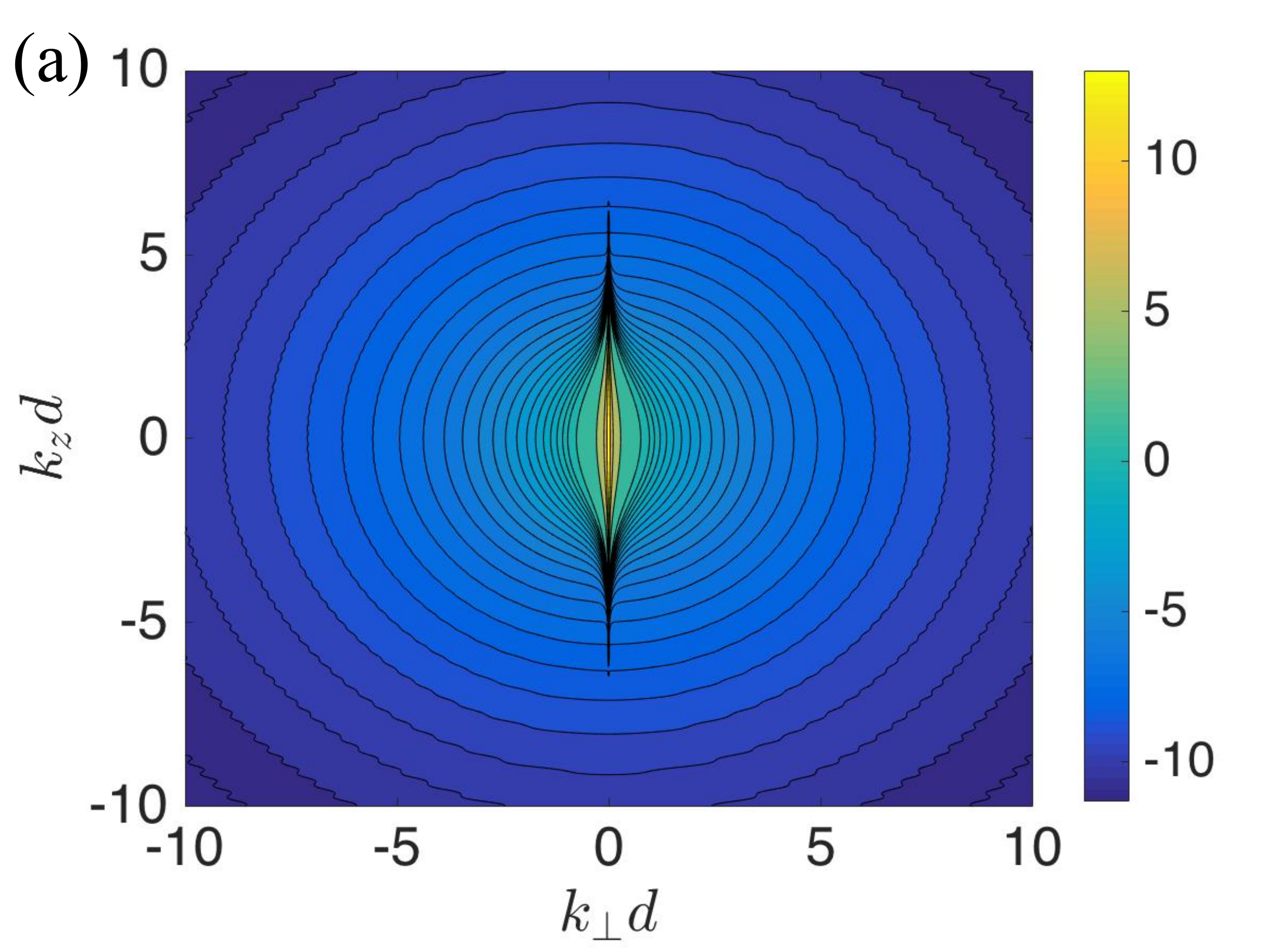}}
	{\includegraphics[width=0.238\textwidth]{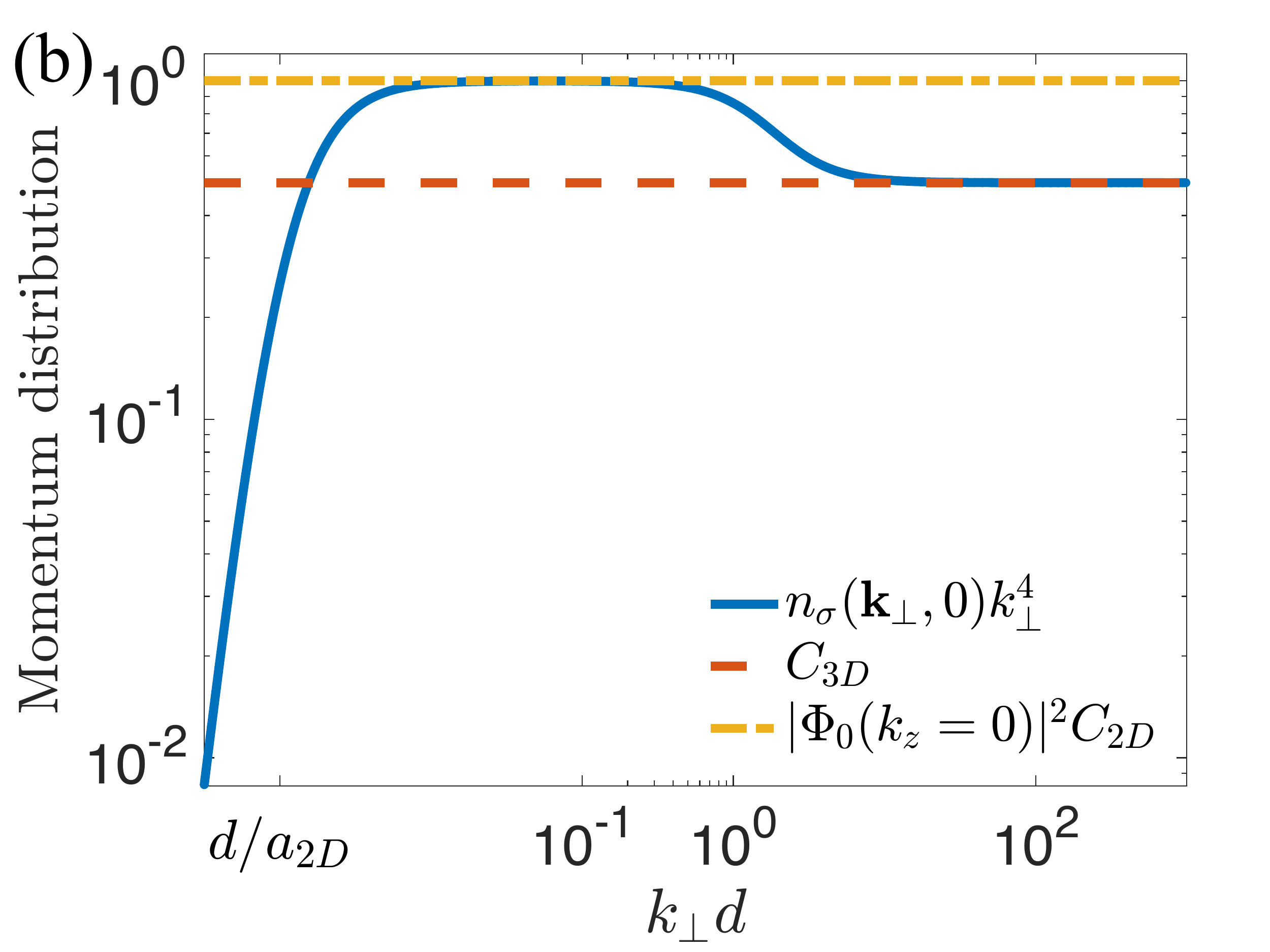}}
	\caption{(a) A contour plot of the exact momentum distribution $\ln(n_\sigma({\bf k}))$ of a two-body system, with $n_\sigma({\bf k})$ in unit of $d^4|\Phi_{0} (k_z=0)|^2 C_{2D}$. The total number of vibration levels considered is $N=300$, and $a_{2D}=1000d$. (b) Scaled momentum $n_\sigma({\bf k}_\bot,0)k_\bot^4$. It is determined by $C_{2D}$ and $C_{3D}$ in the regime $a_{2D}^{-1}\ll k_\bot\ll d^{-1}$ and $k_z\gg d^{-1}$, respectively.  }\label{Fig3}
\end{figure}

Figure \ref{Fig3} shows the numerical results for the momentum distribution of a two-body system. Again, its scaling behaviours capture those of a generic many-body system in the regime, $k\ll k_F$. When $k_F\ll k_\bot \ll d^{-1}$,  where ${\bf k}=({\bf k}_\bot, k_z)$ and $k_\bot =|{\bf k}_\bot|$, we obtain the 2D analogy of Eq. (\ref{nk1Dfull})
\begin{equation}
n_\sigma({\bf k}) \stackrel{ k_F\ll k\ll  d^{-1}}{\xrightarrow{\hspace*{1cm}} } |\Phi_{0}(k_z)|^2\frac{ C_{2D}}{k_\bot^{4}},
\end{equation}
which shows that $n_\sigma({\bf k})$ decays slowly in the $k_z$ direction, a characteristic quasi-2D feature.  We define a 2D momentum distribution in this regime, 
\begin{equation}
n_{2D}^\sigma({\bf k}_\bot)=\int_{-\infty}^{\infty}  \frac{dk_z}{2\pi} n_\sigma({\bf k}) \stackrel{k_F\ll k_\bot \ll d^{-1}}{\xrightarrow{\hspace*{1cm}} } \frac{C_{2D}}{k_\bot^4}.
\label{2Dnk}
\end{equation}
where
\begin{equation}
C_{2D}=(2\pi)^2 N_{\uparrow}N_{\downarrow}\int d^3 {\bf R}_{12} \Big|\int d\epsilon_q G({\bf R}_{12},E-\epsilon_q)\Big|^2.\label{C2D}
\end{equation}
By considering the asymptotic behavior of $\phi({\bf r};q)$ at $\rho \ll d$ and $z=0$, one can also obtain that
\begin{equation}
\phi({\bm \rho},0;\epsilon_q)  \stackrel{ \rho \ll d }{\xrightarrow{\hspace*{1cm}} }  \frac{\sqrt{d\sqrt{\pi}  }}{2} \left( \frac{1}{\rho} - \frac{1}{a_{3D}} \right), \label{Asym3D2D}
\end{equation} 
which is consistent with Eq. (\ref{Asym3D}), and \cite{Petrov}
\begin{equation}
a_{2D}=\sqrt{\frac{2\pi}{\tau}} d \exp{\left(-\frac{\sqrt{\pi}}{2} \frac{d}{a_{3D}} - \gamma \right)}, \label{sl32}
\end{equation}
where $\tau=0.915\cdots$ and $\gamma$ is the Euler's constant,  $\cot{\eta_{2D}}=\frac{2}{\pi} \ln{\left(qa_{2D}e^\gamma/2\right)}$, $G_{3D}({\bf R}_{12};E-\epsilon_q)= \sqrt{d\sqrt{\pi}/4} G({\bf R}_{12};E-\epsilon_q)$. Thus, when $r\ll d$ or equivalently, $k\gg d^{-1}$ , the system is 3D-like, as shown in figure \ref{Fig3}. $n_\sigma({\bf k})$ becomes isotropic and is governed by $C_{3D}$. Compare Eq. (\ref{C3D}) with Eq. (\ref{C2D}), it is clear that Eq. (\ref{C23}) holds. We can also see that
\begin{equation}
n_\sigma({\bf k})k^4\big|_{k\gg d^{-1}}=\sqrt{\pi d^2} n_{2D}^\sigma({\bf k}_\bot){k_\bot^4}\big|_{k_F\ll k_\bot\ll d^{-1}}.
\end{equation}
Similar to the discussions in quasi-1D cases, we find out that the adiabatic relation, 
\begin{equation}
\frac{dE}{d\ln{a_{2D}}}=\frac{\hbar^2C_{2D}}{2\pi M}. \label{ab2D}
\end{equation}
which was originally derived for strictly 2D systems \cite{V3}, still holds for quasi-2D traps. By taking Eq. (\ref{sl32}) and Eq. (\ref{C23}) into Eq. (\ref{ab2D}), it recovers the 3D adiabatic relation in Eq. (\ref{ab3D}).

In conclusion, we have shown an exact relation between $C_{3D}$ and $C_{1D}$ ($C_{2D}$) in quasi-1D (quasi-2D) traps, which correlates not only physical quantities at different length or momentum scales but also universal relations in different dimensions.  We hope that our work will provide physicists a new angle to explore the dimension crossover, and inspire more studies of the central role of contacts in many-body quantum phenomena of quantum gases and related systems.

This work is supported by Hong Kong Research Grants Council/General Research Fund (Grant No. 14306714) and startup funds from Purdue University.

\end{document}